\documentclass[11pt]{article}
\usepackage[T1]{fontenc}
\usepackage{amsfonts}
\usepackage[english]{babel}
\usepackage{a4wide,times}
 
 \usepackage[latin1]{inputenc}
 \usepackage{amssymb}
\usepackage{graphics}
\usepackage{epsfig}
\usepackage{amsbsy}
 \usepackage{verbatim}
\usepackage{color}
\usepackage{mathrsfs}
\usepackage{makeidx}
\usepackage{amsmath}
\usepackage{amsthm}
\usepackage[square,numbers]{natbib}

\newtheorem{thm}{Theorem}[section]
\newtheorem{prop}{Proposition}[section]

\newtheorem{lem}{Lemma}[section]
\newtheorem{rem}{Remark}[section]

\newtheorem*{defin}{Definition}
 \newtheorem*{pre*}{Preuve}

\numberwithin{equation}{section}
\date{}

\title{Pricing of barrier  options  by marginal functional quantization}
\author{ Abass SAGNA   \footnote{ E-mail: abass.sagna@gmail.com. This research is supported by the ``Chaire Risque de Cr\'{e}dit'' of the French Banking Federation.} \\ \\   Laboratoire d'Analyse et de probabilit\'es de \\ l'Universit\'e d'Evry Val d'Essonne  \& \\ ENSIIE}

\begin{document}  
\maketitle

\begin{abstract}
This paper is devoted to the pricing of Barrier  options  by optimal  quadratic quantization method. From a known useful  representation of the premium of barrier options one deduces an algorithm similar to one used to estimate nonlinear filter  using  quadratic optimal functional quantization. Some numerical tests are fulfilled  in the  Black-Scholes model and in a local volatility model and a comparison to the  so called  Brownian Bridge method is also done.  
\end{abstract}

\section{Introduction}
 
 Consider a fixed time horizon  $T$,  which will be typically the maturity of the option in a financial model, and  let $(\Omega,\mathcal{F},\mathbb{P})$ be a probability space  (modeling the randomness of the market) with a filtration $\mathcal{F} = \{\mathcal{F}_t , 1\leq t \leq T\}$ satisfying the usual requirements. 
  The probability $\mathbb{P}$ is supposed to be the probability  in the 'real world' in opposite to the risk neutral probability.

 Consider that the stock price process $(X_t)_{t \in [0.T]}$ satisfies the following  time homogenous   stochastic differential equation (SDE) 
 
 \begin{equation}  \label{EquationSDEGen}
 d X_t = b(X_t) dt + \sigma(X_t) dW_t,   \qquad X_0 = x \in \mathbb{R},
\end{equation}
where $(W_t)_{t \in [0,T]}$ denotes a one-dimensional Brownian motion defined on the probability space $(\Omega,\mathcal{F},\mathbb{P})$;  $b : \mathbb{R} \rightarrow \mathbb{R}$ and  $\sigma : \mathbb{R} \rightarrow \mathbb{R}$
     are continuous functions  satisfying the global Lipschitz and linear growth conditions:
  \begin{equation}  \label{Lipsc_cond}
  \vert  b(x) - b(y) \vert  + \vert \sigma(x) - \sigma(y) \vert  \leq C \vert x-y \vert
  \end{equation}
  and 
  \begin{equation} \label{lineargrowth_cond}
  \vert  b(x) \vert + \vert  \sigma(x) \vert  \leq C (1+ \vert x \vert),
  \end{equation}
for every $t \in [0,T]$ and for every $x,y \in \mathbb{R}$.  The filtration considered  here is the natural  filtration of the brownian  motion completed  by the $\mathbb{P}$-null sets.

It is known that under the above assumptions on the coefficients of the diffusion   there exists a unique strong solution for the SDE (see e.g. \cite{KarShr,Oks}). The uniqueness of the solution is ensured by  the  global Lipschitz assumption  $(\ref{Lipsc_cond})$ whereas  the linear growth  assumption $(\ref{lineargrowth_cond})$  guaranties  that this solution do not explode (see \cite{Oks} for more details). 

 The  first workable model for 'rational' market  pricing of traded options have been proposed by Black-Scholes in $1973$ and extended by Merton in the  same year.  In the Black-Scholes model the economics consists on two assets: the stock price with dynamics as the previous  SDE  with  $b(t,x) := \mu x$ and $\sigma(t,x):= \sigma x$, and a zero-coupon bound of constant interest rate $r$ and maturity $T$.

  Moreover, we know that under arbitrage free and completeness assumptions, the discounted price at time $t$ of any European contingent claim is uniquely determined and is the  expectation, under a probability  $\widetilde{\mathbb{P}}$ called risk neutral probability, of its  discounted payoff  (a functional of the price process $(X_t)_{t \in [0,T]}$ which  may depend on all the trajectory of the process), given all the information available up to time $t$. If $V_t$ is the value of the option a time $t$ and if  $h$ denotes the payoff  at the maturity, then 
 $$ V_t = e^{-r(T-t)} \mathbb{E} (h \vert \mathcal{F}_t),$$
where $\mathbb{E}$ is the expectation under  $\widetilde{\mathbb{P}}$, so that the price at time $0$ is 
\begin{equation} \label{EqTruePrice}
 V_0= e^{-rT} \mathbb{E} (h).
 \end{equation}

Our aim in this work is to estimate  such an expectation for a class of path-dependent payoffs: barrier options, by optimal quantization method.  We consider  here a class of exotic  options whose payoff depend on both the value  of the underlying asset at the maturity and its maximum or its minimum over $[0,T]$. This means, payoffs $h$ of the form
$$ h = F(X_T,\sup_{t \in [0,T]} X_t)  \qquad \textrm{or } \quad h = F (X_T,\inf_{t \in [0,T]} X_t).$$
 
 When the payoff can be decomposed as
 $$  h = \varphi(X_T) \mbox{\bf{1}}_{\{ \sup_{t \in [0.T]} X_t \in I \}} \qquad \textrm{or }  \varphi(X_T) \mbox{\bf{1}}_{\{ \inf_{t \in [0.T]} X_t \in I \}}$$
 where $I$ is an unbounded interval of $\mathbb{R}$, one speaks about barrier options. This last class is a particular case of payoffs of the form
 $$ h = \varphi(X_T) \mbox{\bf{1}}_{\{ \tau_D(X) >T \}}, $$
 where  $\tau_D(X)$ is the exit time of a domain $D \subset \mathbb{R}^d$ by a $d$-dimensional underlying asset $X = (X^1,\dots,X^d)$.
   
   Here are some useful  definitions.
 \vskip 0.2cm
\begin{defin} {\rm
 The option is said  to be an  up-and-out option if  it knocks out  when the price of its underlying  asset  crosses  a specified  value.  It is said  a down-and-out option if it has barrier below the initial   asset price     and knocks out if the underlying asset price falls below the barrier. 

The payoff  of an up-and-out call expiring at time $T$, with strike price $K$ and up-and-out barrier $L$ is  given by : $$ (X_T-K)^{+} \mbox{\bf{1}}_{ \{ \underset{t \in [0.T]}{\sup} X_{t}  \leq L\}} $$
and the payoff of  a  down-and-out  call barrier option with maturity $T$, strike $K$ and barrier $L$ is  given by
$$ (X_T-K)^{+} \mbox{\bf{1}}_{ \{ \underset{t \in [0.T]}{\inf} X_{t}  \geq L\}} . $$
The payoff of put options are defined similarly with $(K-X_T)^{+}$ in place of $(X_T-K)^{+}$.
}
\end{defin}

Note that  closed formulas are available for the price of such options  in the Black-Scholes framework, see \cite{ConVis}. But  this no longer holds when we  move out from  the Black-Scholes framework.  So that we are led to estimate the prices by some numerical procedures.  One of the used methods is the regular Brownian bridge method (see e.g. \cite{Bal}). It provides approximation formulae of the price of barrier options using diffusion bridge methods. This leads to useful forms to approximate these prices from recursive  formulas (already pointed out in \cite{Sag}) similar to an algorithm used in \cite{PagPha} to estimate nonlinear filter by optimal quantization method. One difference of our setting with respect to the one of \cite{PagPha} is that our algorithm involves non-regular  functions. Furthermore, if we consider local volatility model in the previous setting, one way of processing the algorithm is to use Lloyd's algorithm (or stochastic algorithms) to compute the optimal grids and the  transition probabilities.  But, because of the irregularity of functions appearing in our context, one must increase the grid sizes of the marginal quantization of the price process to obtain good approximations of the prices. It is clear that this will be very time consuming to use Lloyd's algorithm to compute grids sizes,  and, this also depends to the parameters of the model.  Moreover,   the marginal quantized process is not a Markov chain and, for numerics, it is forced to satisfy the Markov property.

  In this work, we propose a procedure based on  (quadratic) marginal  functional quantization method. It consists first in considering the ordinary differential equation (ODE)  resulting to  the substitution of  the Brownian motion appearing in the dynamics of the price  process (\ref{EquationSDEGen}) by one quadratic quantization  of  the  Brownian motion. Then, constructing some ``good'' marginal  quantization of the price process based on the solution of the previous ODE's, we show how  to estimate the premium of barrier options from a recursive formula similar to an algorithm used to estimate nonlinear filter using optimal quantization method. Note that by construction,  the marginal  quantized discrete process is a Markov chain. Furthermore, because this procedure is based on the quantization of the Brownian motion, it does not depend on model parameters and price estimates are obtained in few seconds (at most in 6 seconds and sometimes instantaneously, for considered examples).   Numerical simulations are performed in the Black-Scholes model and in the local volatility model called a pseudo CEV model.  A comparison with the Regular Brownian Bridge method show that the former method may some times be faster and competitive with respect to the last one.
 
   The paper is organized as follow. Since in a general setting, the estimation of the prices requires paths discretization of the process, we will recall in Section \ref{SectEulerScheme} the Euler scheme and some relevant convergence rate. Then, we will see in Section \ref{Secpriceestimates}  how to derive the price estimates from (continuous) Euler scheme. The obtained formulas are well known and are  moreover in a useful form to apply  an algorithm similar to that  used in nonlinear filtering estimation via optimal quantization. The algorithm and the relevant error are given in Section \ref{pureoptimquant}. This algorithm involves the marginal quantization of the stock price process and, in Section  \ref{Secfunctionalquant}, we show how to construct such a process from a basic construction of functional quantization of  a diffusion process. We end by some numerical experiments where we compare our method with the regular Brownian bridge method in the Black-Scholes model and in the pseudo CEV model.

 \section{Euler Scheme} \label{SectEulerScheme}
 Consider a one-dimensional Brownian diffusion process $(X_t)_{t \in [0,T]}$, solution of the following stochastic differential equation 
 \begin{equation} \label{EqDSE}
 dX_t = b(X_t) dt + \sigma(X_t) dW_t, \qquad X_0 = x \in \mathbb{R}
 \end{equation}
 where $b :\mathbb{R} \rightarrow \mathbb{R}$, $\sigma : \mathbb{R} \rightarrow \mathbb{R}$ are continuous functions  satisfying  conditions (\ref{Lipsc_cond}), (\ref{lineargrowth_cond}) and  $(W_t)_{t \in [0,T]}$ denotes a one-dimensional Brownian motion defined on  $(\Omega,\mathcal{F},\mathbb{P})$.
 
 Let  us divide the set $[0,T]$ into $n$ subsets  of  length $T/n$ and set for every $k=0,\dots,n,\ t_k = \frac{kT}{n}.$ The stepwise constant Euler scheme is defined by    
 \begin{equation}  \label{EqEulerDiscScheme}
 \tilde{X}_{t_{k+1}} = \tilde{X}_{t_k}   + b(\tilde{X}_{t_k}) \  \frac{T}{n}  + \sigma(\tilde{X}_{t_k}) \sqrt{\frac{T}{n}} Z_{k+1}, \quad \tilde{X}_0 =x, \ k=0, \dots,n-1
 \end{equation}
 where $(Z_k)_{1\leq k \leq n}$ is a sequence of $i.i.d$ random variables distributed as $\mathcal{N}(0;1)$.
 
 For every  $t \in [0,T]$, set    $\underline{t} = t_k $ if  $t  \in [t_k,t_{k+1}), \quad k=0,\dots,n-1.$ 
  A natural extension of the discrete Euler scheme  is the continuous Euler scheme defined for every $t \in  [0,T]$ by
 $$ \bar{X}_t  =  \bar{X}_{\underline{t}} + b(\bar{X}_{\underline{t}}) (t-\underline{t}) + \sigma(\bar{X}_{\underline{t}}) (W_t-W_{\underline{t}}) , \quad \bar{X}_0 = x$$ 
 which satisfies  the  SDE
 $$  \bar{X}_t = x  + \int_0^t b(\bar{X}_{\underbar{s}}) ds + \int_0^t \sigma(\bar{X}_{\underbar{s}}) d W_s. $$
The above paths discretization  methods generate some  errors  which estimates are given in the  following results (see e.g \cite{}).

\vskip 0.2cm
\noindent $\rhd$ {\em Strong error rate.} Assume $b$ and $\sigma$ satisfy for every $\alpha \in (0,1)$,
\begin{equation}
 \forall t \in [0,T], \forall y,z \in \mathbb{R}^d, \quad \vert b(s,y) - b(t,z) \vert \leq C (\vert t-s \vert^{\alpha} + \vert y-z \vert) .
\end{equation}
Then, 

\noindent $(a)$  for every $p>0$, for every  $n \geq 1$,
$$ \Vert \sup_{t \in [0,T]} \vert X_t - \bar{X}_t \vert \Vert_p  \leq C_{b,\sigma,p} \ e^{T C_{b,\sigma,p}} (1+ \vert x \vert) \left(\frac{T}{n} \right)^{\frac{1}{2} \wedge \alpha}; $$

\noindent $(b)$  for every $p>0$, for every $n \geq 1$,
$$ \Vert \sup_{t \in [0,T]} \vert X_t - \tilde{X}_t \vert \Vert_p  \leq C_{b,\sigma,p} \ e^{T C_{b,\sigma,p}} (1+ \vert x \vert) \sqrt{ \frac{\log(n)}{n}} .$$

\vskip 0.2cm 
\noindent $\rhd$ {\em Weak error.}
We recall  some  weak error estimates for  path-dependent options (we refer e.g.  \cite{Gob} for  the proofs). Let  
 $$  \mathbb{D}([0,T],\mathbb{R}^d) := \left\{ \xi :[0,T]  \rightarrow  \mathbb{R}^d, \ \textrm{ c\`adl\`ag} \right\}.$$
If $F:\mathbb{D}([0,T],\mathbb{R}^d) \rightarrow  \mathbb{R}$  is a Lipschitz functional for the sup norm, that is,
$$ \vert  F(\xi) - F(\xi') \vert  \leq C_F \sup_{t \in [0,T]} \vert \xi(t) - \xi'(t) \vert $$ 
then 
\begin{equation}
\big\vert \mathbb{E} F((X_t)_{t \in [0,T]}) -  \mathbb{E} F((\bar{X}_t)_{t \in [0,T]})  \big\vert   \leq \frac{C}{\sqrt{n}}
\end{equation}
and  
\begin{equation}
\big\vert \mathbb{E} F((X_t)_{t \in [0,T]}) -  \mathbb{E} F((\tilde{X}_t)_{t \in [0,T]})  \big\vert   \leq C  \sqrt{\frac{\log n}{n} }.
\end{equation}
  
On the other hand,  if a domain $D$ has a smooth enough boundary, $b, \sigma \in \mathcal{C}^{3}(\mathbb{R})$ and $\sigma$ uniformly elliptic on $D: \ \exists \sigma_0>0, \ \forall x \in \mathbb{R} \ \sigma^2(x)  \geq \sigma_0^2$,  then,  for every bounded  measurable function $f$ satisfying $d({\rm supp}(f),\partial D) \geq 2 \varepsilon >0,$
  $$   \mathbb{E} (f(\bar{X}) \mbox{\bf{1}}_{\{ \tau(\bar{X})>T \}}) -  \mathbb{E} (f(X) \mbox{\bf{1}}_{\{ \tau(X)>T \}})  = C n^{-1} + o(n^{-1})$$
and 
$$   \mathbb{E} (f(\tilde{X}) \mbox{\bf{1}}_{\{ \tau(\tilde{X})>T \}}) -  \mathbb{E} (f(X) \mbox{\bf{1}}_{\{ \tau(X)>T \}})  =  O(n^{-1/2})$$
where  $n$ is the number of discretization steps and  $\tau(Y)$ is the exit time of the process $Y$  from  the  open set  $D$, $i.e$
$$  \tau(Y) = \inf \{t \in [0,T], \ Y_t \in D^c \} .$$
Then the convergence rate is of order $n^{-1}$ for the continuous Euler scheme and of order $n^{-1/2}$ for the discrete one.

\section{Approximation  of knock out option prices using diffusion bridge}  \label{Secpriceestimates}

According to the convergence rate for the continuous Euler scheme we would like  to  estimate  the price of  path-dependent options  by replacing the asset price process $(X_t)_{t \in [0,T]}$ by its continuous Euler process $(\bar{X})_{t \in [0,T]}$ in (\ref{EqTruePrice}). Then, given values of the process $(\bar{X}_t)$ at discrete times $t_k, k=0,\dots,n$, one deduces formulas integrating the useful information  which is the probability that the barrier is not knocked over the time interval $[0,T]$. Remark that this information is lost when replacing $X$ by the discrete Euler process $\tilde{X}$ in (\ref{EqTruePrice}) because we do not known if whether or not the barrier is knocked between time intervals $(t_k,t_{k+1}), k=0, \dots, n-1$. But,  integrating this information in the former case requires the knowledge of the distributions of the maximum and the minimum  of the continuous Euler process $(\bar{X}_t)$ over the time interval $[0,T]$, given its values at  the discrete time observations $t_k$.  
\begin{prop}   We have
\begin{equation}  \label{loi_du_max}
\mathcal{L}(\max_{t \in [0,T]} \bar{X}_t \vert \bar{X}_{t_k} = x_{k}, k=0,\dots,n) = \mathcal{L}(\max_{k=0,\dots,n-1} G^{-1}_{x_{k},x_{k+1}} (U_k) )
 \end{equation}
and
\begin{equation}  \label{loi_du_min}
\mathcal{L}(\min_{t \in [0,T]} \bar{X}_t \vert \bar{X}_{t_k} = x_{k}, k=0,\dots,n) = \mathcal{L}(\min_{k=0,\dots,n-1} F^{-1}_{x_{k},x_{k+1}} (U_k) )
 \end{equation}
where $(U_k)_{k=0,\dots,n-1}$ are $i.i.d$ random variables uniformly  distributed over the unit interval, $G^{-1}_{x,y}$ and $F^{-1}_{x,y}$ are the inverse functions of  the conditional distribution functions  $G_{x,y}$  and $F_{x,y}$ defined by
$$G_{x,y} (u)  = \left (1 -  e^{-2n \frac{(x-u)(y-u)}{ T \sigma^2(x)}} \right) \mbox{\bf{1}}_{\{ u \geq \max (x,y) \}}$$
and
$$F_{x,y} (u)  = 1- \left(1-e^{-2n \frac{(x-u)(y-u)}{ T \sigma^2(x)}}  \right)\mbox{\bf{1}}_{\{u \leq \min (x,y) \}}.$$
\end{prop}
\vskip 0.2cm 
This result is proved using the independence property of the processes $(\bar{X}_{t})_{t \in [t_k,t_{k+1}]}$, for $k=0,\dots,n-1$,  given the $\bar{X}_{t_k}=x_k$, and the knowledge of the distribution of the supremum (and the infimum) of brownian bridge diffusion over time intervals $(t_k,t_{k+1})$, with end points $x_k$ and $x_{k+1}$. 

From the above  proposition we deduce  general formulas  making a connexion between   the expectation of a functional of both  the terminal value $\bar{X}_T$ of the  process $(\bar{X}_t)$ and  its maximum (or the minimum)  over the time interval $[0,T]$.  From now on we make the abuse of notation $\bar{X}_k:=\bar{X}_{t_k}, \ \forall k \in \{0,\dots,n \}.$

\begin{prop} \label{prop_exp_max_double}
 $(a)$ Let $f$ be a real-valued non negative   function defined on $\mathbb{R}_{+}^2$ such that  $ f(x,\cdot)$  is a nonnegative   function  satisfying    
\begin{equation}  \label{Assump_unif_finite_var}
\underset{x>0}{\sup} \ \mathbb{E} f(x,\max_{t \in [0,T]} \bar{X}_t) <+\infty .
\end{equation}
Then 
\begin{equation} \label{zero_exp_double_formula_max}
\mathbb{E} f(\bar{X}_T,\max_{t \in [0,T]} \bar{X}_t) = \mathbb{E}f(\bar{X}_T,0) + \mathbb{E} \int_0^{+\infty} \left(1-\prod_{k=1}^n G_{\bar{X}_{k-1},\bar{X}_{k}}(z) \right)  d_zf(\bar{X}_T,z).   
\end{equation}
Likewise if $$ \sup_{x >0} \mathbb{E} f(x,\min_{t \in [0,T]} \bar{X}_t) <+\infty)$$   then
\begin{equation}  \label{zero_exp_double_formula_min}
\mathbb{E} f(\bar{X}_T,\min_{t \in [0,T]} \bar{X}_t \vee 0) = \mathbb{E}f(\bar{X}_T,0) + \mathbb{E} \int_0^{+\infty} \left( \prod_{k=1}^n \big(1- F_{\bar{X}_{k-1},\bar{X}_{k}}(z) \big) \right) d_zf(\bar{X}_T,z).   
\end{equation}
 \vskip 0.2cm
\noindent $(b)$  If  furthermore  $ f_{\infty}(x) := \underset{y \rightarrow +\infty}{\lim} f(x,y) <+\infty$  for every $x>0$.  Then 
\begin{equation}  \label{infty_exp_double_formula_max}
\mathbb{E} f(\bar{X}_T,\max_{t \in [0,T]} \bar{X}_t) = \mathbb{E}f_{\infty}(\bar{X}_T) - \mathbb{E} \int_0^{+\infty} \left(\prod_{k=1}^n G_{\bar{X}_{k-1},\bar{X}_{k}}(z) \right)  d_zf(\bar{X}_T,z)   
\end{equation}
and 
\begin{equation}   \label{infty_exp_double_formula_min}
\mathbb{E} f(\bar{X}_T,\min_{t \in [0,T]} \bar{X}_t \vee 0 ) = \mathbb{E}f_{\infty}(\bar{X}_T) - \mathbb{E} \int_0^{+\infty} \left(1-\prod_{k=1}^n \big(1-F_{\bar{X}_{k-1},\bar{X}_{k}}(z) \big)\right)  d_zf(\bar{X}_T,z).   
\end{equation}
\end{prop}

\noindent This proposition follows from the following lemma.

\begin{lem}  \label{lem_exp_max} 
$(a)$ \  Let $Z$ be a positive random variable and let $g$ be a  nonnegative function with  finite variation (on compact sets) such that 
\begin{equation} \label{assump_alem_exp_max}
 \mathbb{E} \left( \int_{]0,Z]} \vert dg \vert  \right)  <+\infty.
\end{equation}
   Then 
 \begin{equation}
  \mathbb{E} g(Z) = g(0) + \int_{(0,+\infty)} \mathbb{P}(Z\geq z) \ dg(z).
  \end{equation}

\vskip 0.2cm
\noindent  $(b)$ If furthermore  $g_{\infty}:= \underset{x \rightarrow +\infty}{\lim} g(x) < +\infty$ then        

 \begin{equation}
  \mathbb{E} g(Z) = g_{\infty} -  \int_{(0,+\infty)} \mathbb{P}(Z < z) \ dg(z).
  \end{equation}

\end{lem}
\vskip 0.3cm

\begin{proof}[$\textbf{Proof}$]  $(a)$ \  We have
$$ g(Z) = g(0) + \int_{]0,Z]} dg(u). $$
It follows that
\begin{eqnarray*}
 \mathbb{E} g(Z)  & = & g(0)  + \mathbb{E}  \int_{]0,Z]} dg(u)  \\
 & = &  g(0) +  \int_{(0,+\infty)}  \mathbb{P}(Z \geq z) \ dg(z),
 \end{eqnarray*}
the last inequality  coming from Fubini's theorem; which can be applied owing to assumption $(\ref{assump_alem_exp_max})$.
\vskip 0.2cm
\noindent $(b)$ \ Just use the fact that  $\mathbb{P}(Z \geq u) = 1 -   \mathbb{P}(Z < u). $
\end{proof}

Now we are in position to prove Proposition $\ref{prop_exp_max_double}$.
\begin{proof}[$\textbf{Proof of Proposition $\ref{prop_exp_max_double}$}$]
$(b)$  One deduces from $(\ref{loi_du_max})$   that 
$$ \mathbb{E} \big( f(x_n,\max_{t \in [0,T]} \bar{X}_t )\big \vert \bar{X}_{k} = x_k,k=0,\dots,n \big)  = \mathbb{E} \big(f(x_n,\max_{0 \leq k \leq n-1} G^{-1}_{x_k,x_{k+1}}(U_k) ) \big) $$
where $G^{-1}_{x,y}$  and the $U_k$ are defined like in $(\ref{loi_du_max})$. Then, applying  Lemma $\ref{lem_exp_max} \ (b)$  to  the function  $g(z)=f(x_n,z)$   gives  
\begin{eqnarray*}
\mathbb{E} \big(f(x_n,\max_{0 \leq k \leq n-1} G^{-1}_{x_k,x_{k+1}}(U_k) ) \big )
&  = & f_{\infty}(x_n) - \int_0^{+\infty} \mathbb{P}(\max_{0 \leq k \leq n-1} G^{-1}_{x_k,x_{k+1}}(U_k) \leq z ) d_zf(x_n,z) \\
& = &  f_{\infty}(x_n)  -  \int_0^{+\infty} \left(\prod_{k=0}^{n-1}\mathbb{P}(U_k \leq G_{x_k,x_{k+1}}(z)) \right)d_zf(x_n,z)  \\
& = &  f_{\infty}(x_n)  -  \int_0^{+\infty} \left(\prod_{k=0}^{n-1}G_{x_k,x_{k+1}}(z)) \right) d_zf(x_n,z) .
\end{eqnarray*}
Consequently 
\begin{eqnarray*}
 \mathbb{E} f(\bar{X}_T,\max_{t \in [0,T]} \bar{X}_t)  
 & =  & \mathbb{E} \big(\mathbb{E} \big( f(\bar{X}_T,\max_{t \in [0,T]} \bar{X}_t )\big \vert \bar{X}_{k} = x_k,k=0,\dots,n \big) \big) \\
 &  = & \mathbb{E}f_{\infty}(\bar{X}_T) - \mathbb{E} \int_0^{+\infty} \left(\prod_{k=1}^n G_{\bar{X}_{k-1},\bar{X}_{k}}(z) \right)  d_zf(\bar{X}_T,z).   
 \end{eqnarray*}
 The formula relative to the minimum is proved likewise  by using  $(\ref{loi_du_min})$  in place of  $(\ref{loi_du_max})$. 
 \vskip 0.2cm
 \noindent $(a)$ is proved like $(b)$ by using Lemma $\ref{lem_exp_max} \ (a)$ instead of Lemma $\ref{lem_exp_max} \ (b)$.
\end{proof}

Proposition $\ref{prop_exp_max_double}$ allows  us to rewrite the  estimates of the  premiums  of  some  usual exotic options (in particular barrier options)  in a useful form in view of  the optimal quantization approximation method as well as  of Monte Carlo simulation methods.  Let us  mention that the following  representations of the price estimates of Barrier options are well known  even if the computational  method used here to derive them is a little  different.
\begin{prop} \label{Propformulas}
 Let  $f(x) = (x-K)^{+}$ and $g(x) = (K-x)^{+}$. 
\vskip 0.2cm
\noindent  $(a)$ The price of  an  up-and-out put option  expiring at time  $T$ with strike $K$ and up-and-out barrier  $L$  is  estimated   by  
\begin{equation}  \label{UpOutPut}
\bar P_{UB}:= e^{-rT} \mathbb{E} \big((K-\bar{X}_T)^{+} \mbox{\bf{1}}_{\{ \underset{t \in [0,T]}{\sup} \bar{X}_t \leq L \}} \big) =  e^{-rT}\mathbb{E} \left( g(\bar{X}_T) \prod_{k=1}^{n} G_{\bar{X}_{k-1},\bar{X}_k}(L)  \right).
\end{equation}

\noindent   $(b)$ The  price of  an up-and-out call option  expiring at time  $T$ with strike $K$ and up-and-out barrier  $L$  can be approximated by
\begin{equation}  \label{eqUpOutCall}
\bar C_{UB}:= e^{-rT} \mathbb{E} \big((\bar{X}_T-K)^{+} \mbox{\bf{1}}_{\{ \underset{t \in [0,T]}{\sup} \bar{X}_t \leq L \}} \big) =  e^{-rT}\mathbb{E} \left( f(\bar{X}_T) \prod_{k=1}^{n} G_{\bar{X}_{k-1},\bar{X}_k}(L)  \right).
\end{equation}

\noindent  $(c)$ The price of  an  down-and-out put option  expiring at time  $T$ with strike $K$ and down-and-out barrier  $L$   can be approximated by
\begin{equation}  \label{DownOutPut}
\bar P_{OB}:= e^{-rT} \mathbb{E} \big((K-\bar{X}_T)^{+} \mbox{\bf{1}}_{\{ \underset{t \in [0,T]}{\inf} \bar{X}_t \geq L \}} \big) =  e^{-rT}\mathbb{E} \left( g(\bar{X}_T) \prod_{k=1}^{n} \big(1- F_{\bar{X}_{k-1},\bar{X}_k}(L) \big) \right).
\end{equation}

\noindent  $(d)$ The  price of  an  down-and-out  call option  expiring at time  $T$ with strike $K$ and up-and-out barrier  $L$  is approximated  by the following formula :
\begin{equation}  \label{DownOutCall}
\bar C_{OB}:= e^{-rT} \mathbb{E} \big((\bar{X}_T-K)^{+} \mbox{\bf{1}}_{\{ \underset{t \in [0,T]}{\inf} \bar{X}_t \geq L \}} \big) =  e^{-rT}\mathbb{E} \left( f(\bar{X}_T) \prod_{k=1}^{n}\big(1- F_{\bar{X}_{k-1},\bar{X}_k}(L)  \big) \right).
\end{equation}
\end{prop}
\vskip 0.2cm

Note that  the right hand side of Equations  $(\ref{UpOutPut})$, $(\ref{eqUpOutCall})$, $(\ref{DownOutPut})$, $(\ref{DownOutCall})$  are   obtained by  re-conditioning.  Then,  it follows from Jensen inequality that the corresponding variances are smaller than the variances induced  by the left hand side of the same equations.

\begin{proof}[$\textbf{Proof}$]
  $(a)$   Let $f(x) = (x-K)^{+} $, $g(z) =  \mbox{\bf{1}}_{\{ z \leq L \}} $  and set $h(x,z) =f(x) g(z)$. Then it follows from  $(\ref{infty_exp_double_formula_max})$  that  
 $$ \mathbb{E} \big((\bar{X}_T-K)^{+} \mbox{\bf{1}}_{\{ \underset{t \in [0,T]}{\sup} \bar{X}_t \leq L \}} \big)  = \mathbb{E} h_{\infty} (\bar{X}_T) -\mathbb{E} \int_0^{+\infty} \left(f(\bar{X}_T) \prod_{k=1}^n G_{\bar{X}_{k-1},\bar{X}_{k}}(z) \right)  dg(z) .$$
 Now  $\ \forall x \geq 0, \ h_{\infty}(x) = 0  $ and $dg(z) = -\delta_{L}(z)$. Then 
 $$  C_{UB} =  e^{-rT}\mathbb{E} \left( f(\bar{X}_T) \prod_{k=1}^{n} G_{\bar{X}_{k-1},\bar{X}_k}(L)  \right). $$
 The items $(b)$, $(c)$, $(d)$  are proved in the same way as $(a)$.
\end{proof}
In the next section, we show how to estimate the previous prices by optimal quantization. We will give first an approximating algorithm and then, the induced error. Since this algorithm involves the quantization of the price process and its transition probabilities, we will point out how to construct a functional quantization of the  price process  and how  to estimate its transition probabilities.

\section{Estimation of the prices by marginal quantization} \label{pureoptimquant}
The aim of this section is to propose an algorithm based  on optimal quantization to compute the path dependent options  pointed out in Section $\ref{Secpriceestimates}$.  To this end we will approximate by optimal quantizations some expressions of the form  
\begin{equation}
V:= \mathbb{E} \left( f(\bar{X}_n) \prod_{k=1}^{n} g_{k}(\bar{X}_{k-1},\bar{X}_{k}) \right)
\end{equation}
where  $f$ is  a bounded measurable function  on $\mathbb{R}^d$ taking values on $\mathbb{R}$  and $g_k(\cdot,\cdot)$ a measurable function on $\mathbb{R}^d \times \mathbb{R}^d $ which may  depend on some real parameters like for  Barrier options where it depends also on the  barrier. 
\subsection{The algorithm}  \label{SectAlgorithm}
We will mainly refer  to \cite{PagPha}, where  numerical solving of nonlinear filtering  with discrete-time observation have been performed by optimal quantization methods. The only  change is that in our  setting we will drop the dependance on the noisy observations  ($i.e$ the $Y_k$'s following the notations in \cite{PagPha}) because  our problem of interest here  is not a filtering problem.

We  define  for any $k=1,\dots,n$,  the bounded transition kernel $H_k$ by 
\begin{equation} \label{def_kernel_H}
 H_{k} f(x) = \mathbb{E} \left( f(\bar{X}_{k}) g_k({x,\bar{X}_{k}} ) \vert  \bar{X}_{k-1} =x \right) = \int f(y) g_k(x,y) P_k(x,dy) 
 \end{equation}
where $P_k(x,\cdot) = \mathcal{L}(\bar{X}_k = \cdot \vert \bar{X}_{k-1} =x)$.  For convenience, we set
\begin{equation} \label{def_kernel_H0}
 H_0 f(x) = \mathbb{E}(f(\bar{X}_0))  = \int f(x) \mu(dx) .
 \end{equation}
\noindent 
Now for any $k \in \{1,\dots,n \}$ set
$$ \pi_k f= \mathbb{E} \big( f(\bar{X}_k) \prod_{i=1}^{k}  g_k(\bar{X}_{i-1},\bar{X}_{i}) \big).$$
We have 
\begin{eqnarray*}
\pi_k f & =  &  \mathbb{E} \left(  \mathbb{E} \big( f(\bar{X}_k) \prod_{i=1}^{k}  g_i(\bar{X}_{i-1},\bar{X}_{i})  \vert \mathcal{F}_{t_{k-1}} \big)  \right)  \\
&  = &   \mathbb{E} \left(  \mathbb{E} \big( f(\bar{X}_k)  g_k(\bar{X}_{k-1},\bar{X}_{k}) \vert \bar{X}_{k-1} \big)   \prod_{i=1}^{k-1}  g_i(\bar{X}_{i-1},\bar{X}_{i})   \right)  \\
& = &   \mathbb{E} \left(  H_k (f( \bar{X}_{k-1}))    \prod_{i=1}^{k-1}  g_i(\bar{X}_{i-1},\bar{X}_{i})  \right)
\end{eqnarray*}
It follows  that  
\begin{equation}   \label{eq_recurrence}
\pi_k f = \pi_{k-1} H_k f , \quad k=1,\dots,n 
\end{equation}
so that  
\begin{equation}  \label{EqEstimPi}
 V = \pi_n f = (H_0  \circ H_1 \circ \dots \circ H_n) f .
\end{equation}

\vskip  0.3cm

 Then, to estimate  $V$ we  need to approximate $\pi_n$.  At this step, suppose that  we have access to the quantization $(\widehat{X})_{t_{k}}$  of the price process  over the time steps $t_k, k=0,\dots,n$ on grids $\Gamma_k =\{ x_k^1,\dots,x_k^{N_k} \}$ of sizes $N_k$, $k=0,\dots,n$ (see further on for facts about quantization).    


Owing to equation $(\ref{eq_recurrence})$  our aim is to estimate the price  using an approximation of  the probability transition $P_k(x_k,dx_{k+1})$  of $\bar{X}_{k+1}$ given $\bar{X}_k$. These  probability transitions are approximated  by the probability transition matrix  $\hat{p}_k := (\hat{p}_k^{ij})$  of $\widehat{X}_{k+1}$ given   $\widehat{X}_k$:
\begin{equation} \label{transition_weights}
\hat{p}_k^{ij} = \mathbb{P} (\widehat{X}_k = x_k^j  \vert \widehat{X}_{k-1} = x_{k-1}^i);  \ i=1,\dots,N_{k-1}, \ j=1,\dots,N_k.
\end{equation}

Then, following Equation $(\ref{def_kernel_H})$,  we estimate the transition kernel matrix  $H_k$ by the quantized  transition kernel $\widehat{H}_k$  given by    
$$ \widehat{H}_k  = \sum_{j=1}^{N_k} \widehat{H}_k^{ij} \delta_{x_{k-1}^i}, \quad k=1,\dots,n$$
where
\begin{equation}  \label{EqDefHIJ}
 \widehat{H}_k^{ij} =   g_k(x_{k-1}^i,x_k^j)  \hat{p}_k^{ij}, \quad i=1,\dots,N_{k-1}, \ j=1,\dots,N_k .
 \end{equation}
For $k=0$, we set   (owing to $(\ref{def_kernel_H0})$ and to the fact that $X_{0} = x_{0}$ is not random) 
$$ \widehat{H}_0 =  \delta_{x_0}.$$
  We  finally approximate  $\pi_n$ by 
  
 \begin{equation}   \label{approxim_hatpi}
 \widehat{\pi}_n = \widehat{H}_0 \circ \widehat{H}_1 \circ \dots \circ \widehat{H}_n  ;
 \end{equation} 
which  in turn can be  computed by the forward induction

\begin{equation} 
  \widehat{\pi}_0 = \widehat{H}_0, \qquad   \widehat{\pi}_k = \widehat{\pi}_{k-1} \widehat{H}_k, \quad \qquad  k=1,\dots,n.
 \end{equation}
It follows that  the price  $V=\pi_n f $  may be estimated by  summery 
$$ \widehat{V} := \widehat{\pi}_n f .$$

From the previous approach, we deduce the following estimations for options of interest  using  optimal functional  quantization method. Set in this scope  $f(x)  := (x-K)^{+}$  and $g(x) := (K-x)^{+}$.
\vskip 0.3cm
$\rhd$ {\em Up-and-out options.} According to the forgoing we estimate the price of an up-and-out put option by  
$$  \widehat{P}_{UB}  := e^{-rT}  \widehat{\pi}_n g $$
and the price of  up-and-out call option is approximated by
$$  \widehat{C}_{UB}  := e^{-rT}  \widehat{\pi}_n f $$
where  $ \widehat{\pi}_n $  is  defined  as in $(\ref{approxim_hatpi})$   with  the associated  transition kernel 
$$ \widehat{H}_k^{ij} =   G_{x_{k-1}^i,x_k^j }(L) \hat{p}_k^{ij}, \quad i=1,\dots,N_{k-1}; \ j=1,\dots,N_k.$$
\vskip 0.2cm

$\rhd$  {\em Down-and-out options.}   The down-and-out  put option's price is estimated by 
$$  \widehat{P}_{OB} := e^{-rT}  \widehat{\pi}_n g $$
and the  price of  down-and-out barrier call option is estimated by
$$  \widehat{C}_{OB} := e^{-rT}  \widehat{\pi}_n f $$
where  for both cases   $ \widehat{\pi}_n $  is  defined  as in $(\ref{approxim_hatpi})$   with  the associated  transition kernel 
$$ \widehat{H}_k^{ij} =   F_{x_{k-1}^i,x_k^j }(L) \hat{p}_k^{ij}, \quad i=1,\dots,N_{k-1}; \ j=1,\dots,N_k.$$
\vskip 0.2cm
\begin{rem}  \label{RemAccelCompational}
 {\rm 
 One  numerical advantage of this algorithm is that $\widehat{\pi}_n$ does not depend on the function  $f$ appearing in (\ref{EqEstimPi}).  Then, once   $\widehat{\pi}_n$  is computed we deduce both the call and the put price approximations.  On the other hand, 
considering  Equation (\ref{EqDefHIJ}) one notices that as soon as $\widehat{X}_{k-1}$  reaches the barrier  (for example, for the up-and-out option: there exists $i_0$ such that $x_{k-1}^{i_0}>L$), then,  $\widehat{H}_k^{ij}=0$ for every $i \geq i_0$.   For numerical computation,  we may  take account of  this fact  to  reduce the computation time.} 
\end{rem}

\subsection{Error analysis}
In order to have some upper bound of the quantization error  estimate of   $\pi f$ we  need the following assumptions $\textbf{(A1)}$ and $\textbf{(A2)}$ :  
\vskip 0.3cm
\noindent $\textbf{(A1)}$ \ The transition operator $P_k(x,dy)$ of $X_k$ given $X_{k-1}, \ k=1,\dots,n$ are Lipschitz.
\vskip 0.3cm
Recall that a probability transition $P$ on $\mathbb{R}^d$ is ${\rm C}$-Lipschitz  (with ${\rm C}>0$) if for any Lipschitz function $f$ on $\mathbb{R}^d$ with ratio $[f]_{Lip}$, $Pf$ is Lipschitz with ratio $[Pf]_{Lip} \leq {\rm C} [f]_{Lip}$.  Then, one may define the Lipschitz ratio $[P]_{Lip}$ by 

\begin{equation*}
[P]_{Lip} = \sup \big\{ \frac{[Pf]_{Lip}}{[f]_{Lip}}, f  \textrm{ a  nonzero Lipschitz function }  \big\} < +\infty . 
\end{equation*}
Then if  the transition  operators  $P_k(x,dy),\ k=1,\dots,n$   are Lipschitz,   it follows  that
$$ [P]_{Lip} := \max_{k=1,\dots,n}[P_k]_{Lip} <+\infty. $$
\vskip 0.3cm
\noindent $\textbf{(A2)}$ \ It   consists on the following two assumptions.
\begin{itemize}
\item[$(i)$]  For every $k=1,\dots,n$, the functions  $g_{k}(\cdot,\cdot)$ are bounded on $\mathbb{R}^d \times \mathbb{R}^d$ and we set 
  $$ {\rm K}_g := \max_{k=1,\dots,n} \Vert  g_{k}  \Vert_{\infty} $$
  \item[$(ii)$]  For every $k=1,\dots,n$, there  exist  two  constants $[g^1_{k}]_{Lip}$ and $[g^2_{k}]_{Lip}$ so that for every $x,x',\widehat{x},\widehat{x}' \in \mathbb{R}^d$,
  $$  \vert g_{k}(x,x') - g_{k}(\widehat{x},\widehat{x}') \vert  \leq [g^1_{k}]_{Lip}\ \vert x-\widehat{x} \vert + [g^2_{k}]_{Lip} \ \vert x'-\widehat{x}' \vert .$$
\end{itemize}

\begin{thm}
Under Assumptions $\textbf{(A1)}$ and  $ \textbf{(A2)}$ we have for every  bounded Lipschitz  continuous function $f$ on $\mathbb{R}^d$ and for every  $p \geq 1$,

\begin{equation} \label{Eqerrorbound}
 \vert   \pi_n f - \widehat{\pi}_n f \vert   \leq \sum_{k=0}^n {\rm C}_k^n(f,p) \ \Vert X_k - \widehat{X}_k  \Vert_p
\end{equation}
with 
$$ {\rm C}_k^n(f,p)  =  (2-\delta_{2,p}) \ {\rm K}^{k}_g  \ [u_k]_{Lip} +  {\rm K}^{n-1}_g \ \Vert f \Vert_{\infty} ([g^1_{k+1}]_{Lip} + [g^2_{k+1}]_{Lip})  . $$
\end{thm}
\vskip 0.3cm

\begin{proof}[$\textbf{Proof}$] The proof follows from  the proof of Theorem $3.1$  in  \cite{PagPha} by dropping the dependency on the  noisy observations $(y_1,\dots,y_n)$ following  the notations of the authors. 
\end{proof}
Now, let us come back to the construction of the quantized price process  $(\widehat{X}_k)_{k=0, \dots,n}$. We show in the next section how to construct this process after making a short background on product functional  quantization of gaussian processes, in particular, of brownian motion. 
\section{Marginal functional  quantization of the price process}  \label{Secfunctionalquant}
Before dealing with the construction of the marginal functional quantization of the price process, we  make some background on functional quantization of gaussian processes.
\subsection{A brief overview on functional  product  quantization of gaussian processes}
We remind  first   some basic notions  about optimal vector quantization.  It is a process of approximating a continuous range of values or a very large set of discrete values by a relatively small set of discrete values. Rigorously speaking, the $L^r$-optimal quantization  problem at level $n$ for  a  $\mathbb{R}^d$-valued random vector $X$  lying in   $L^r(\Omega,\mathcal{A},\mathbb{P})$ consists in finding  the best approximation of  $X$  by a  Borel function of $X$ taking  at most  $n$ values. This problem can be reads as
\begin{eqnarray}
 e_{n,r}(X)  & = & \inf{\{ \Vert X - \widehat{X}^{\alpha} \Vert_r, \alpha \subset \mathbb{R}^d,  \textrm{ card}(\alpha) \leq n \}}  \nonumber \\
  & = &  \inf_{ \substack{\alpha \subset \mathbb{R}^d \\ \textrm{card}(\alpha) \leq n}} \left(\int_{\mathbb{R}^d} d(x,\alpha)^r dP(x) \right)^{1/r}. \label{er.quantchap3}
 \end{eqnarray}
where $ \widehat{X}^{\alpha} = \sum_{a \in \alpha} a \mbox{\bf{1}}_{\{X \in C_a(\alpha)\}}$  is the quantization of $X$  on the grid $\alpha$ and $(C_a(\alpha))_{a \in \alpha}$  corresponds to a Voronoi tessellation  of  $\mathbb{R}^d$ (with respect to a norm $\vert \cdot \vert$ on $\mathbb{R}^d$), that is, a Borel partition of $\mathbb{R}^d$ satisfying for every $a \in \alpha$, $$ C_a(\alpha) \subset \{ x \in \mathbb{R}^d : \vert x-a \vert = \min_{b \in \alpha} \vert x-b \vert \}.$$

The quantity  $ e_{n,r}(X)$  is called the $L^{r}$-mean quantization error. This error decreases to zero  at a $n^{-1/d}$-rate as the size $n$ of the codebook $\alpha$ goes to infinity. This convergence rate has been investigated in \cite{BucWis} and  \cite{Zad}  for absolutely continuous probability measures under the  quadratic norm on $\mathbb{R}^d$ and studied in great details  in \cite{GraLus} under an arbitrary norm on  $\mathbb{R}^d$  for  absolutely continuous measures  and some singular measures. Very recently, optimal  vector quantization has become a promising tool in Numerical Probability owing  to  its ability  to approximate  either expectations  or more significantly  conditional expectations from some  cubature formulas.  This faculty to approximate conditional expectations is the crucial  property  used to solve some problems emerging in finance as optimal stopping problems (pricing and hedging American style options, see   \cite{BalPag,PagPri},  stochastic control problems  (see \cite{CorPhaRun,PagPhaPri}) for portfolio  management, nonlinear filtering problems  (see \cite{PagPha,PhaRunSel} and \cite{CalSag} for an application to credit risk).   
   
  A rigorous extension of  optimal vector quantization to functional quantization is done in \cite{LusPag} where the vector quantization problem is transposed to random variables taking values in an infinite dimensional Hilbert space, in particular, to stochastic processes $(X_{t})_{t \in [0,1]}$ viewed as random variables with values in  $L^{2}([0,1],dt)$. Many others works have been done in this direction as e.g. \cite{DerFehMatSch}. From the numerical point of view, it is pointed out   in \cite{PagPri}  how a Gaussian process  can be quantized using  Karhunen-Lo\`eve  product quantization based on product quantization of Gaussian random variables  coming from  the  Karhunen-Lo\`eve expansion of the  given Gaussian process.  A closed formula for the distribution of the quantization of the Gaussian process (in particular for Brownian motion) is derived and some applications  has been successfully  performed in  Finance, namely,  in the pricing  of vanilla and Asian call options in Heston model.
    
   To recall some basic results about functional quantization  suppose that  $(H,(\cdot \vert \cdot)_{H})$  is a separable Hilbert space and let $X: (\Omega,\mathcal A, \mathbb P) \mapsto H$  be square integrable $H$-valued  random vector with distribution $\mathbb P_{X}$ defined on $(H,\mathcal Bor(H))$ where  $\mathcal Bor(H)$ stands for  the 
Borel $\sigma$-field.  Let $\Vert  \cdot  \Vert$  denotes the  $L^{2}_{H}(\Omega,\mathbb P)$-norm defined by $\Vert  X \Vert_{2}^{2} = \mathbb E (\vert X \vert_{H}^{2})$.

Let $x :=\{x_{1},\dots,x_{n}\}  \in H^{n}$ be an $n$-quantizer  and let  $\widehat{X}^{x}$ be the quantization of $X$ on the grid $x$ defined previously, where   the Voronoi tessellation $(C_{i}(x))_{1 \leq i \leq n}$  induced by $x$ satisfies  for every $i  \in \{1,\dots,n \}$,
$$ C_i(x) \subset \{ y \in H : \vert x_{i}-y \vert_{H} = \min_{1 \leq j \leq n} \vert x_{j}-y \vert_{H} \}.$$
The quadratic quantization problem consists  of finding an optimal quantizer $x \in H^{n}$ (if any), means, an  $n$-quantizer  which minimizes the quantization error $\Vert X -\widehat{X}^{x} \Vert_{2}$ over $H^{n}$. From the numerical integration viewpoint, finding an optimal quantization may be a difficult problem and we are sometimes let to find some 'good' quantization $\widehat{X}^{x}$  which is close to $X$ in distribution, so that  for every Borel function $F: H \mapsto \mathbb R$, we can approximate $\mathbb E F(X)$ by  
\begin{equation}  \label{EqCubature}
 \mathbb E F (\widehat{X}^{x}) = \sum_{i=1}^{n}  F(x_{i}) \mathbb P_{X} (C_{i}(x)).
 \end{equation}
 Then  if we have access to both the $n$-quantizer $x=\{x_{1}, \dots,x_{n} \}$ and the distribution associated to $\widehat{X}^{x}$, $\big(\mathbb P_{X} (C_{i}(x)) \big)_{1 \leq i \leq n}$, the estimation of $\mathbb E F(X)$ using  Equation (\ref{EqCubature}) is straightforward. The induced error depends on the regularity of the functional $F$ and here is some error bounds.  Suppose that  $X \in L^{2}_{H}(\Omega,\mathbb P)$ and let $F$ be a Borel functional defined on $H$.
\begin{enumerate}
\item If $F$ is Lipschitz continuous with Lipschitz constant $[F]_{{\rm Lip}}$ then for every $n$-quantizer $x$,
$$  \vert  \mathbb E F(X) - \mathbb E F(\widehat{X}^{x}) \vert \leq [F]_{{\rm Lip}} \Vert  X - \widehat{X}^{x} \Vert_{2}$$
so that if $(x_{n})_{n \geq 1}$ is a sequence of quantizers satisfying $\underset{n \rightarrow \infty}{\lim} \Vert  X - \widehat{X}^{x_{n}} \Vert_{2} = 0$,  then  $\widehat{X}^{x_{n}}$  converge in distribution to $X$.
\item If $F$ is differentiable on $H$  with an $\theta$-H\"older  differential $DF$, $\theta \in (0,1]$, then for every optimal $n$-quantizer $x$,
$$  \vert  \mathbb E F(X) - \mathbb E F(\widehat{X}^{x}) \vert \leq [DF]_{\theta} \Vert  X - \widehat{X}^{x} \Vert_{2}^{1+\theta}.$$
\end{enumerate}
Some others bound are available (we refer to \cite{PagPri} for more detail). Now let us say how to get some 'good' quantizers for Gaussian processes  to  make sense  the previous errors bounds.  We  consider here a centered one-dimensional $L^{2}_{T}:=L^{2}([0,T],dt)$-valued  Gaussian process $X$ satisfying   
$$ \mathbb E \vert X \vert_{L^{2}_{T}}^2  =  \int_{0}^{T} \mathbb E (X^{2}_{s}) ds< + \infty. $$
The process $X$ admits the following representation in the Karhunen-Lo\`eve  basis  (see e.g. \cite{LusPag})
\begin{equation}   \label{EqProdQuant}
  X(\omega)   \stackrel{L^{2}_{T}}{ = }  \sum_{k \geq 1} \sqrt{\lambda_{k}} \xi_{k}(\omega) e_{k}^{X} \qquad \mathbb P(d \omega)-{\rm a.s.} 
  \end{equation}
where  the sequence $(\xi_{k})_{k \geq 1}$ defined for every $k \geq 1$ by
\begin{equation}  \label{EqK-Lexpansion}
 \xi_{k} = \frac{ (X \vert e_{k}^{X} )}{\sqrt{{\rm Var}((X \vert e_{k}^{X} ))}} 
 \end{equation}
is a sequence of $i.i.d$  $\mathcal N(0;1)$-distributed random variables. Owing to  the expansion (\ref{EqK-Lexpansion}), a natural way to produce a functional product quantization of a Gaussian process in $L^{2}_{T}$ of size at most $N$  is to use a product quantizer of the form
\begin{equation}  \label{EqProdQuantizers}
  \widehat{X}_{t}^{(d_{N})} =  \sum_{k = 1}^{L} \sqrt{\lambda_{k}} \ \widehat{\xi}^{x^{(N_{k})}}_{k} e_{k}^{X}(t) 
\end{equation}
where $ \widehat{\xi}^{x^{(N_{k})}}_{k} $  is an optimal $N_{k}$-quantization of $\xi_{k}$ and $d_{N}:= N_{1}\times \dots \times N_{L} \leq N$, with $N_1,\dots,N_L \geq 2$. An quadratic optimal product $N$-quantizers also noted  $\widehat{X}_{t}^{(d_{N})}$ is obtained by solving the  optimization problem (see \cite{LusPag} for more detail):
\begin{equation} \label{EquatProbBit}
 \min \big\{ \Vert  X - \widehat{X}^{(d_{N})}   \Vert_{2}, \ d_N=N_{1} \times \dots \times N_{L} \leq N; N_1,\dots,N_L \geq 2;  L \geq 1  \big\}.
 \end{equation}
We suppose from now on that  the previous optimization problem can be solved, at least  numerically,  and  that the optimal $L$-tuple still be denoted by $N_1,\dots,N_L$.
Then, numerical computation of  a Gaussian process $X$ is possible as soon as we have numerical access to the eigensystem  $(e_{n}^{X},\lambda_{n})$,  which, for the Brownian motion $(W_{t})_{t \in [0,T]}$, has a closed formula:
$$ e_{k}^{W}(t) := \sqrt{\frac{2}{T}} \sin \Big( \pi (k-1/2) \frac{t}{T}\Big)   \quad \textrm{and} \quad \lambda_{k} := \Big(\frac{T}{\pi(k-1/2)} \Big)^{2}, \ k \geq 1. $$
So,  the  one-dimensional quadratic optimal product quantizer $\alpha^N$,  at level  $N$, of the Brownian motion $(W_{t})_{t \in [0,T]}$, is defined by   
\begin{equation*}  
\alpha^N_{i_1,\dots,i_L}(t)  =  \sqrt{\frac{2}{T}} \sum_{k=1}^{L} \frac{T}{\pi (k-1/2)} \sin\Big( \pi (k -1/2) \frac{t}{T}\Big)  x_{i_{k}}^{(N_{k})}, \  1 \leq i_{k} \leq N_{k}, \ 1 \leq k \leq L, 
\end{equation*}
where  $x^{(N_{k})} =\{x_{1}^{N_{k}}, \dots, x_{N_{k}}^{N_{k}}  \}$  is the optimal quantization of the $\mathcal{N}(0;1)$ of  size $N_{k}$ and   $d_{N} = \prod_{k=1}^{L}N_{k}$ is an optimal integer solving Problem  (\ref{EquatProbBit}) (with respect to the Brownian motion).  Remark that for every $t \in [0,T]$, the marginal quantizer $\alpha^N_{i_1,\dots,i_L}(t)$ is of size $d_N$.  For numerics, a whole package of product $N$-quantizers of the standard Brownian motion  are available at  {\tt  www.quantize.} {\tt maths-fi.com}.
We move now to the construction of the quantized  price  process. 
\subsection{Marginal functional quantization and transition probabilities}
Recall that the continuous Euler price process  evolves following the SDE
 \begin{equation}  \label{EqPriceProcIntSecMarg}
 d X_t = b(X_{\underline{t}}) dt + \sigma(X_{\underline{t}})  dW_t,   \qquad X_0 = x.
\end{equation}
Let  $(\alpha^{N})_{N \geq  1}$, with for every $N \geq 1$, $\alpha^N(t)=\{\alpha^N_1(t),\dots,\alpha^N_{d_N}(t) \}$,  be a sequence of  optimal product $N$-quantizers of the Brownian motion  and let 
\[
\widehat{W}^{(d_N)}_{t} = \sum_{m=1}^{d_N} \alpha^N_m(t) \mbox{\bf{1}}_{\{ W_t \in C_m(\alpha^N(t)) \}},   \qquad t \in [0,T], 
\]
be the marginal functional quantization of the Brownian motion.  It is known that  the sequence $(\alpha^N)_{N \geq 1}$ is  rate-optimal,  means, 
\[
\Vert  W - \widehat{W}^{\alpha^N} \Vert_2 = O \big((\log N)^{-1/2} \big).
\]
 Consider  the sequence $x^N = (x_m^N)_{m =1,\dots,d_N}$, $N\geq 1$ of solutions of the ODE's 
\begin{equation}  \label{EqDefEDO}
 x_m^N (t) = x +\int_0^t \big[b(x_m^N(\underline{s})) -\frac{1}{2} \sigma\sigma'(x_m^N(\underline{s}))\big] ds + \int_0^t \sigma(x_m^N(\underline{s})) d \alpha_m^N(s), \quad m=1,\dots,d_N
 \end{equation}
$\underline{t} = t_k $ if  $t  \in [t_k,t_{k+1}),  \ k=0,\dots,n-1$, and define  the marginal functional  quantization  of $\bar{X}_t$ over the grid  $x^N(t) = \{x_1^N(t),\dots,x_{d_N}^{N}(t) \}$ by
\begin{equation}  \label{EqDefQuantPriceProc}
 \widehat{X}_t^N = \sum_{m=1}^{d_N} x_m^N(t) \mbox{\bf{1}}_{\{\bar{X}_t  \in C_m(x^N(t)) \}},
 \end{equation}
 so that Equation (\ref{EqDefEDO}) can be written as
 \begin{equation*}  
 \widehat{X}_t^N  = x +\int_0^t \big[b( \widehat{X}^N_{\underline{s}}) -\frac{1}{2} \sigma \sigma'(\widehat{X}^N_{\underline{s}}) \big] ds + \int_0^t \sigma( \widehat{X}^N_{\underline{s}}) d \widehat{W}_s^{(d_N)}.
 \end{equation*}

 Recall that the process $(\bar{X}_{t_k})$ is a Markov chain. Then, since by construction  $\sigma(\bar{X}_{t_{k}}, k=0,\dots,n) = \sigma(\widehat{X}^{N}_{t_k},k=0,\dots,n)$, the discrete process $(\widehat{X}^N_{t_k})_{k=0,\dots,n}$ is a Markov chain. 
On the other hand, since the additional  term $\frac{1}{2}\sigma \sigma'$ appears in the ODE (this correction term can be dropped by considering the stochastic integral  in  (\ref{EqPriceProcIntSecMarg}) in the sense of Stratonovich integral for $L^p_{L^2_T}(\Omega,\mathbb P)$ convergence investigation tools, see \cite{LusPag}), we must make the supplementary  assumption that  $\sigma$ is continuously differentiable with bounded derivative  to guaranty the existence and the uniqueness of the solution.  Now, given the quantization process $(\widehat{X}^N)$, to  complete the estimation of  the price of barrier options following the introduced  algorithm in Section  \ref{SectAlgorithm}, it suffice to compute the transition probabilities appearing in (\ref{transition_weights}). Notice  that  all our  grids $x^N(t_k)$   are of size $d_N=N_1 \times \dots N_L$. To define correctly the Voronoi cell associated  to the grids $x^N(t_k)$ we will consider that for every time step $t_k$, $x^N(t_k) = \{x_1^{N}(t_k), \dots,x_{d_N}^{N}(t_k) \}$ is a  descendent ordered set. The computation of the transition probabilities will be made differently according to the following two situations.
\vskip 0.2cm
\noindent  $\rhd$ {\em The cumulative distribution function  $F(\cdot;x)$ of  the  conditional law of  $X_t$ given $X_s=x$ is known for every $s \leq t$.} For example, this is the case in  the Black Scholes model where $F(\cdot;x)$ is the cumulative distribution function of the lognormal distribution.  In this case,  since following Equation (\ref{EqDefQuantPriceProc}),  $\sigma(\bar{X}_{t_{k}}, k=0,\dots,n) = \sigma(\widehat{X}^{N}_{t_k},k=0,\dots,n)$,  the probabilities  are estimated by
\begin{equation} \label{EqApproxProba}
 \hat{p}_{k}^{ij} \approx  F\big(x^N_{j+}(t_{k});x^N_{i}(t_{k-1}) \big) - F\big(x^N_{j-}(t_{k});x^N_{i}(t_{k-1}) \big),
 \end{equation}
with for every $k=0,\dots,n-1$,

\[
    \left \{ \begin{array}{ll}
 x^N_{j+}(t_{k}) := \frac{x^N_{j}(t_{k}) + x^N_{j+1}(t_{k})  }{2}; \ x^N_{j-}(t_{k}) := \frac{x^N_{j}(t_{k}) + x^N_{j-1}(t_{k})  }{2}; \ j=1,\dots,d_{N}-1; \\
 \\
x^N_{1-}(t_k)=0; \ x^N_{d_N^{+}}(t_k) =+\infty.
 \end{array}  
 \right.
 \]

In fact, we have  for every $k=0,\dots,n-1$,
\begin{eqnarray*}
 \hat{p}_{k}^{ij}  &  = & P\big(\bar{X} _{k} \in C(x^N(t_{k})) \big\vert  \bar{X} _{k-1} \in C(x^N(t_{k-1})) \big)  \\
 & \approx & P\big(X_{k+1} \in C_j(x^N(t_{k})) \big\vert  X _{k-1} \in C_i(x^N(t_{k-1}))\big)  \\
 &  = & P\big(X_{k} \leq x^N_{j+}(t_{k})  \big\vert  X _{k-1} \in C_i(x^N(t_{k-1})) \big)  - P\big(X_{k} \leq x^N_{j-}(t_{k})  \big\vert  X _{k-1} \in C_i(x^N(t_{k-1})) \big).
\end{eqnarray*}
Afterward, we have for every $z \geq 0$,
\begin{eqnarray*}
P\big(X_{k} \leq z \vert  X _{k-1} \in C_i(x^N(t_{k-1})) \big)  & = & \frac{P\big(X_{k} \leq z;  X _{k-1} \in C_i(x^N(t_{k-1})) \big) }{ P\big(X _{k-1} \in C_i(x^N(t_{k-1})) \big)},
\end{eqnarray*}
and considering the numerator in the right hand side of the previous equation we have
\setlength\arraycolsep{0.2pt}
\begin{eqnarray}  \label{EqavantApproxquant}
P\big(X_{k} \leq z ;  X _{k-1} \in C_i(x^N(t_{k-1})) \big) & = & \int_{-\infty}^{z} \left(\int_{ C_i(x^N(t_{k-1}))} P(X_k \in dx \vert X_{k-1} = y) dP_{X_{k-1}}(y) \right) dx \nonumber \\
& = & \int_{ C_i(x^N(t_{k-1}))} F(z;y) dP_{X_{k-1}}(y) \\
& \approx &  F(z;x^N_{i}(t_{k-1})) P\big(X _{k-1} \in C_i(x^N(t_{k-1})) \big).  \nonumber
\end{eqnarray}
The last quantity  is the approximation of the right hand side of (\ref{EqavantApproxquant})  by optimal quantization with one grid's point, considering that $\{x_i^N(t_{k-1}) \}$ is  the quantizer of size one of  the random variable $\bar{X}_{t_{k-1}}$ over the Voronoi cell $C_i((x^N(t_{k-1}))$.


\vskip 0.2cm
\noindent $\rhd$  {\em The cumulative distribution function  $F(\cdot,x)$ of  the  conditional law of  $X_t$ given $X_s=x$ is unknown.} In this case, considering the (discrete) Euler Scheme of the price process (see  (\ref{EqEulerDiscScheme}))  we estimate $F$ by  the cumulative distribution function  $\tilde{F}$ of the  $\mathcal N (m_{k};\sigma^{2}_{k})$  with 
$$ m_{k} = \widehat{X}_{k-1} + b \big(t_{k},\widehat{X}_{k-1} \big) \frac{T}{n};  \quad  \sigma^{2}_{k} =  \sigma^{2} \big(t_{k},\widehat{X}_{k-1} \big) \frac{T}{n}, $$
so that for every $i,j =1, \dots, d_{N}$,
\begin{equation} \label{EqEstTransProba}
  \hat{p}_k^{ij} \approx  \tilde{F}\big(x^N_{j+}(t_{k});x^N_{i}(t_{k-1}) \big) - \tilde{F}\big(x^N_{j-}(t_{k});x^N_{i}(t_{k-1}) \big)
    \end{equation}
where the  $x^N_{j+}(t_{k})$  and $x^N_{j-}(t_{k})$   are defined as previously.

Notice that since the error  bound of the filter estimate  in (\ref{Eqerrorbound}) involves the  marginal quantization error: $\Vert  \bar{X}_{t_k} - \widehat{X}^{N}_{t_k} \Vert_2,$ one must deduce this error from the above construction. We know that the  sequence of non-Voronoi quantization  $(\tilde{X}^{x^{N}} )_{N \geq 1}$ defined for every $N \geq 1$ by  
  \[
  \tilde{X}^{x^N}_t = \sum_{m=1}^{d_N} x^N_m(t) \mbox{\bf{1}}_{\{ W \in C_m(\alpha^N) \}}
  \]
 is rate-optimal in  $L^p_{L_T^2}(\Omega,\mathbb P)$ for $p \in [1,2)$: $\Vert  \vert  X - \tilde{X}^{x^N}\vert_{L^2_T }\Vert_p = O\big( (\log N)^{-1/2}\big)$ (see \cite{LusPag1}). One theoretical challenge will  be to compute the convergence rate for the marginal functional quantization error.

\section{Numerical illustration}
We deal with numerical experiments by considering  an Up-and-out  call option  in the Black-Scholes model  and a local volatility model  already  considered in \cite{LemPag} and called pseudo CEV model. Recall that in the Black-Scholes framework the stock price process  $(X_t)$ is modeled by the following SDE (under the risk neutral probability $\tilde{\mathbb{P}}$)
\begin{equation}
 dX_t = r X_t dt + \sigma X_t dW_t , \quad X_{0} =x_{0}
 \end{equation}
where $r$ is the interest rate, $\sigma$ the volatility and $W$ a brownian motion under $\tilde{\mathbb{P}}$.  For the pseudo CEV model, the dynamics of  the stock price  process is ruled  by the following SDE (under the risk neutral probability)
\begin{equation}
dX_t = rX_t dt + \vartheta X_t^{\delta} \frac{X_t}{\sqrt{1+X_t^2}} dW_t, \quad X_{0} = x_{0}
\end{equation}
for some $\delta \in (0,1)$ and $\vartheta \in (0,\underline{\vartheta}], \underline{\vartheta}>0$. The parameter $r$ still be the interest rate and $\sigma(x) := \vartheta \frac{x^{\delta}}{\sqrt{1+x^2}}$ corresponds to the local volatility function.  We notice that  for a fixed $\delta \in (0,1)$, if the initial value of the stock process  $X_0$ is  large enough then the pseudo CEV model is  very close to  the CEV model 
$$  dX'_t  = X'_t (rdt + \vartheta (X'_t)^{\delta-1}dW_t ).$$
In particular, for numerical tests we will consider that  $\vartheta \thickapprox   \sigma X_0^{1- \delta}$ where $\sigma$ denotes the regular volatility. The only "aim of the really" rough calibration  is just to deal with reasonable values to obtain prices close to those given by  the Black-Scholes model.   For all the experiments  we set the interest rate $r$ equal to $0.15$. The maturity  is set to $T=1$, the initial value of the stock process  $x_{0} =100$ and $\delta =0.5$ (in the local volatility model).  For numerics, the solution of the  ODE given in (\ref{EqDefEDO}) is approximated by a sixth order Runge-Kutta scheme and  marginal quantizations are of size $d_N = 966$ (corresponding to the optimal decomposition $N_1=23, N_2 = 7, N_3=3, N_4 = 2$, for the problem (\ref{EquatProbBit}), see \cite{PagPri}).

 In the Black-Scholes model we compare the prices computed from the  quantization of the  continuous Euler  process using  (\ref{EqDefEDO}) and (\ref{EqDefQuantPriceProc}) (which prices are  referred by QEP prices)  with the  regular Brownian Bridge method (RBB prices), given the true prices  obtained from a semi-closed formula  available  in \cite{ConVis}.  The regular Brownian Bridge (RBB) method is some efficient  method to compute expressions like
  $$\mathbb{E} f(\bar{X}_T,\sup_{t \in [0,T]} \bar{X}_t)  \quad \textrm{or } \ \mathbb{E} f(\bar{X}_T,\inf_{t \in [0,T]} \bar{X}_t), $$ 
 based on Proposition  $\ref{loi_du_max}$ and consisting (for example for the  estimation of $\mathbb{E} f(\bar{X}_T,\sup_{t \in [0,T]} \bar{X}_t) $) in the following steps : 

\vskip 0.2cm
\noindent   Set $S^f = 0$.

\noindent {\tt for } $m=1$ {\tt to} $M$
\begin{itemize}
\item Simulate a path of the discrete time Euler scheme $(\bar{X}^{(m)})$ and set $x_k = \bar{X}_{t_k}^{(m)}, k=0,\dots,n.$
\item Simulate ${\rm \Gamma}^{(m)} := \underset{0 \leq k \leq n}{\max} (G_{x_k,x_{k+1}})^{-1}(U_k^{(m)})$, where $(U_k^{(m)})_{1 \leq k \leq n}$ are iid with $\mathcal{U}([0,1])$-distribution.
\item compute $f(\bar{X}_T^{(m)},{\rm \Gamma}^{(m)})$.
\item Compute $S_{m}^f := f(\bar{X}_T^{(m)},{\rm \Gamma}^{(m)}) + S_{m-1}^f.  $
\end{itemize}

\noindent {\tt end}. $(m)$ 
\vskip 0.2cm
\noindent with 
$$ (G_{x,y})^{-1}(1-u) = \frac{1}{2}\big( x+y + \sqrt{(x-y)^2-2T\sigma^2(x) \log(u)/n} \big),  \ u \in (0,1).$$
 Then for large enough $M$, 
$$  \mathbb{E} f(\bar{X}_T,\sup_{t \in [0,T]}  \bar{X}_t)  \thickapprox \frac{S_M^f}{M}.$$
 
  The numerical results are depicted in Table \ref{tabPrixUpOutCallSig0p07} and Table \ref{tabPrixUpOutCallSig0p1} for varying values of the barrier $L$ and the volatility $\sigma$.  The number  $M$ of Monte Carlo simulations is set to $10^6$.   For the quantization methods, the computation times  (QEP c.t.) varies from  $0$ to $3$ seconds when $n=10$ and, from $1$ to $6$ seconds when $n=20$, increasing with the barrier as pointed out in  Remark  \ref{EqDefHIJ}. However, for the RBB, the computation time is  of $2$ seconds when $n=10$ and of $5$  seconds when $n=20$. The obtained results show that the quantization method may sometimes  be competitive with respect to the regular Brownian Bridge method, specially  for small number of time discretization steps $n$.
   

\begin{table}[htbp]
 \begin{center}
  \begin{tabular}{l*{6}{l}}
   &      &    & ${\bf \sigma=0.07, n=10}$   &  &  \\
   
  \hline
 L  & {\bf True prices} &  RBB prices & RBB var.   & QEP prices & QEP c. t.   \\
  \hline  
105  & {\bf 0.034} & 0.035 & 0.086     & 0.035 & <1s \\
  110&   {\bf 00.59}  &  00.60 & 2.942    & 00.59  & 1s \\
115  &  {\bf 02.58} & 02.62 & 15.80   & 02.59 & 2s \\
 120 &  {\bf 06.01}   &  06.11    & 33.54    & 06.03  & 2s \\
  125   &  {\bf  09.58} & 09.89  &  41.76 &  09.60  & 2s \\
 130  & {\bf 12.07}  &  12.13 & 43.09& 12.08  & 3s\\
 \hline
  \end{tabular} 
\caption{ \label{tabPrixUpOutCallSig0p07} \small{ Up-and-out Call prices from the quantization method (QEP prices) and the regular brownian bridge method (RBB prices) in the Black-Scholes model for $r =0.15$, $\sigma = 0.07$, $d_{N}=966$, $T=1$, $K=100$, $X_{0}=100$, ${\bf n=10}$ and for varying values of the barrier $L$. QEP c.t. is the quantization method computation time.} }
\end{center}
\end{table}

%

\begin{table}[htbp]
 \begin{center}
  \begin{tabular}{l*{6}{l}}
   &      &     &  ${\bf \sigma=0.07, n=20}$&          &  \\
   
  \hline
 L  & {\bf True prices} &  RBB prices  & RBB var.   & QEP prices  & QEP c. t.   \\
  \hline  
105  & {\bf 0.034} & 0.035 & 0.086  &   0.034 & 1s \\
  110&  {\bf 00.59}  &   00.60 &  2.942    & 00.59 & 3s  \\
115  & {\bf 02.58} & 02.59 & 15.80     & 02.59 & 4s\\
 120 & {\bf 06.01}   & 06.05  &  33.54    & 06.02 & 4s \\
  125   & {\bf  09.58} & 09.64 &  41.76  & 09.59 & 5s  \\
 130  & {\bf 12.07}  &  12.10 & 43.09 &  12.08 & 6s  \\
 \hline
  \end{tabular} 
\caption{ \label{tabPrixUpOutCallSig0p07} \small{ Up-and-out Call prices from the quantization method (QEP prices) and the regular brownian bridge method (RBB prices) in the Black-Scholes model for $r =0.15$, $\sigma = 0.07$, $d_{N}=966$, $T=1$, $K=100$, $X_{0}=100$, ${\bf n=20}$ and for varying values of the barrier $L$. } }
\end{center}
\end{table}

%
%

 \begin{table}[htbp]
 \begin{center}
  \begin{tabular}{l*{5}{l}}
   &          &  ${\bf \sigma=0.1, n=20}$&      &   &  \\
   
  \hline
 L  & {\bf True prices }&  RBB prices & RBB var.  & QEP prices    \\
  \hline  
105  & {\bf 0.029} & 0.029 & 0.067   &  0.029   \\
  110& {\bf  00.42}  & 00.43   &   1.933   &   00.42    \\
115  &  {\bf 01.70} &  01.72 & 10.46   & 01.71   \\
 120 &  {\bf 03.95}  &  03.98   & 26.42   & 03.97      \\
  125   & {\bf  06.70}&   06.76   & 43.82  & 06.72  \\
 130  &  {\bf 09.31}  & 09.38 & 57.19 & 09.34   \\
 \hline
  \end{tabular} 
  
%

\caption{ \label{tabPrixUpOutCallSig0p1} \small{ Up-and-out Call prices from the quantization method (QEP prices) and the regular brownian bridge method (RBB prices) in the Black-Scholes model   for $r =0.15$, $\sigma = 0.1$, $d_{N}=966$, $T=1$, $K=100$, $X_{0}=100$, ${\bf n=20}$, and for varying values of the barrier $L$. } }\end{center}
\end{table}
 

\begin{table}[htbp]
 \begin{center}
  \begin{tabular}{l*{6}{l}}
   &       &     ${\bf \vartheta=0.7, n=20}$&    &  &      \\
   
  \hline
 L   &  {\bf Ref. Price} & RBB price & RBB var. & QEP price  & QEP c. t.    \\
  \hline  
105   & \textbf{0.034} & 0.034 &  0.085  &  0.034  & 2s  \\
 106  & \textbf{0.074} &  0.074 &  0.222   &   0.074 & 2s    \\
  107  & \textbf{00.14} &  00.14&  0.496   &   00.14 & 2s    \\
  110  & \textbf{00.59} &  00.59 &  2.949   &   00.59 & 3s    \\
   111  & \textbf{00.86} &  00.86 &  4.636   &   00.86 & 4s    \\
   112  & \textbf{01.20} &  01.20 &  6.841   &   01.20 & 4s    \\
115  &  \textbf{02.64} &  02.66 & 16.34 & 02.66 &5s \\
 120  &   \textbf{06.25}   &   06.29    & 34.47    & 06.30 & 6s      \\
  125 &  \textbf{09.98} & 10.00  &  41.73 & 10.02 & 7s \\
 130 & \textbf{12.44} &  12.45   &   41.65 & 12.46 & 8s   \\
 \hline
  \end{tabular} 
\caption{ \label{tabPrixUpOutCallVLSig0p07} \small{ Up-and-out Call prices from RBB  and QEP methods in the local volatility model. Model parameters:  $r =0.15$, $\delta=0.5$, $\vartheta = 0.07$, $d_{N}=966$, $T=1$, $K=100$, $X_{0}=100$ and for varying values of the barrier $L$. } }
\end{center}
\end{table}

 \begin{table}[htbp]
 \begin{center}
  \begin{tabular}{l*{5}{l}}
   &     &  ${\bf \vartheta=1.0, n=20}$  &       &   \\
   
  \hline
 L   &  {\bf Ref. Price}  & RBB price & RBB var. & QEP price     \\
  \hline  
105  & {\bf 0.029} & 0.029 & 0.067  &  0.029    \\
106  & {\bf 00.06} & 00.06 & 0.165  &  00.06    \\
107  & {\bf 00.11} & 00.11 & 0.165  &  00.11    \\
  110 & {\bf 00.43} & 00.43   & 1.966    &  00.43     \\
  111  & {\bf 00.61} & 00.61 & 3.022  &  00.61    \\
    112  & {\bf 00.83} & 00.84 & 4.423  &  00.84    \\
115   &{\bf 01.77}  & 01.77& 10.89  & 01.78    \\
 120   &{\bf 04.16}  & 04.20   &   27.81 &  04.20     \\
  125   &{\bf  07.11}  & 07.14   & 45.41   & 07.17   \\
 130   &{\bf  09.87}   & 09.90 & 58.17 & 09.92     \\
 \hline
  \end{tabular} 

\caption{ \label{tabPrixUpOutCallVLSig0p1} \small{ Up-and-out Call prices from RBB and QEP methods in the local volatility model  for $r =0.15$, $\delta=0.5$, $\vartheta = 0.1$, $d_{N}=966$, $T=1$, $K=100$, $X_{0}=100$ and for varying values of the barrier $L$. } }\end{center}
\end{table}

%


 For the local volatility model we compare the QEP prices with the prices obtained from regular Brownian bridge method.  Numerical results are depicted in Tables \ref{tabPrixUpOutCallVLSig0p07} and \ref{tabPrixUpOutCallVLSig0p1}  for different volatilities and for different values of the barrier. Our reference prices are computed from the regular Brownian bridge method with $10^{7}$ Monte Carlo simulations and $100$ times discretization  steps. We remark that  the quantization method is competitive (because it is faster with the same precision) with respect to the regular Brownian bridge method when the barrier is closed to $X_0$. But, the  QEP prices become less precise when the volatility and the barrier increase (for example when $\sigma=1.0$ and $L=130$, where the absolute error is of $2 \%$ with respect to RBB method).  This might be  due to the additional  error coming from the estimation of the transition probabilities by formula (\ref{EqEstTransProba}) since the conditional law  is not known.
 
Notice that, since we have used sixth order Runge-Kutta scheme to approximate  solutions of the ODE (\ref{EqDefEDO}), there is no hope to improve of price approximations for  quantization method by increasing the number $n$ of discretization steps, so that the RBB method will become more competitive when increasing $n$.  To improve the estimations for quantization method we must increase the size of $d_N$. But, this will increase the computation time and the method will become  less faster than the RBB one.

\vskip 1cm

\end{document}